\documentclass[jkps,preprint,fleqn,showpacs,showkeys,linenumbers]{revtex4}
\usepackage{graphicx}
\usepackage{amssymb}
\usepackage{mathtools}
\usepackage{multirow}
\usepackage{units}
\usepackage{siunitx}
\usepackage{bm}
\usepackage{subfig}
\usepackage{caption}
\usepackage[utf8]{inputenc}

\begin{document}
\setcounter{page}{0}

\title[]{Fast Simulation of a Silicon-Pad Detector}
\author{Beomkyu \surname{Kim}}
\author{Jeongsu \surname{Bok}}
\email{jeongsu.bok@cern.ch}
\author{Jaeyoon \surname{Cho}}
\author{Jiyeon \surname{Kwon}}
\author{Hyungjun \surname{Lee}}
\author{Minjung \surname{Kweon}}
\email{minjung.kweon@cern.ch}
\affiliation{Department of Physics, Inha University, Incheon 22212}


\begin{abstract}

Several types of detectors are used to detect charged particles in particle and nuclear physics experiments. Since the semiconductor detector has superior spatial and kinematic resolutions as well as good response time than other types of detectors, it has become one of the most important detectors recently. When charged particles pass through the semiconductor detector, electron-hole pairs are formed inside the detector and move toward the electrode by the electric field inside the detector. At this time, the trajectory and momentum can be determined through the generated current signal.

In this study, we introduce an open-source application named Fast Silicon Device Simulation that is developed for fast simulation of a typical silicon semiconductor detector, such as a p-type pad on an n-type wafer with a reverse-bias voltage. Iterative and multi-grid methods are used to calculate the potential and electric field in the simulation fast. Current signals produced by the simulation are compared with results by Silvaco TCAD and Garfield++ simulations. The simulation program is based on the ROOT that has been developed by CERN.

\end{abstract}



\maketitle

\section{INTRODUCTION}

\begin{figure}[!htb]
\includegraphics[width=0.6\textwidth]{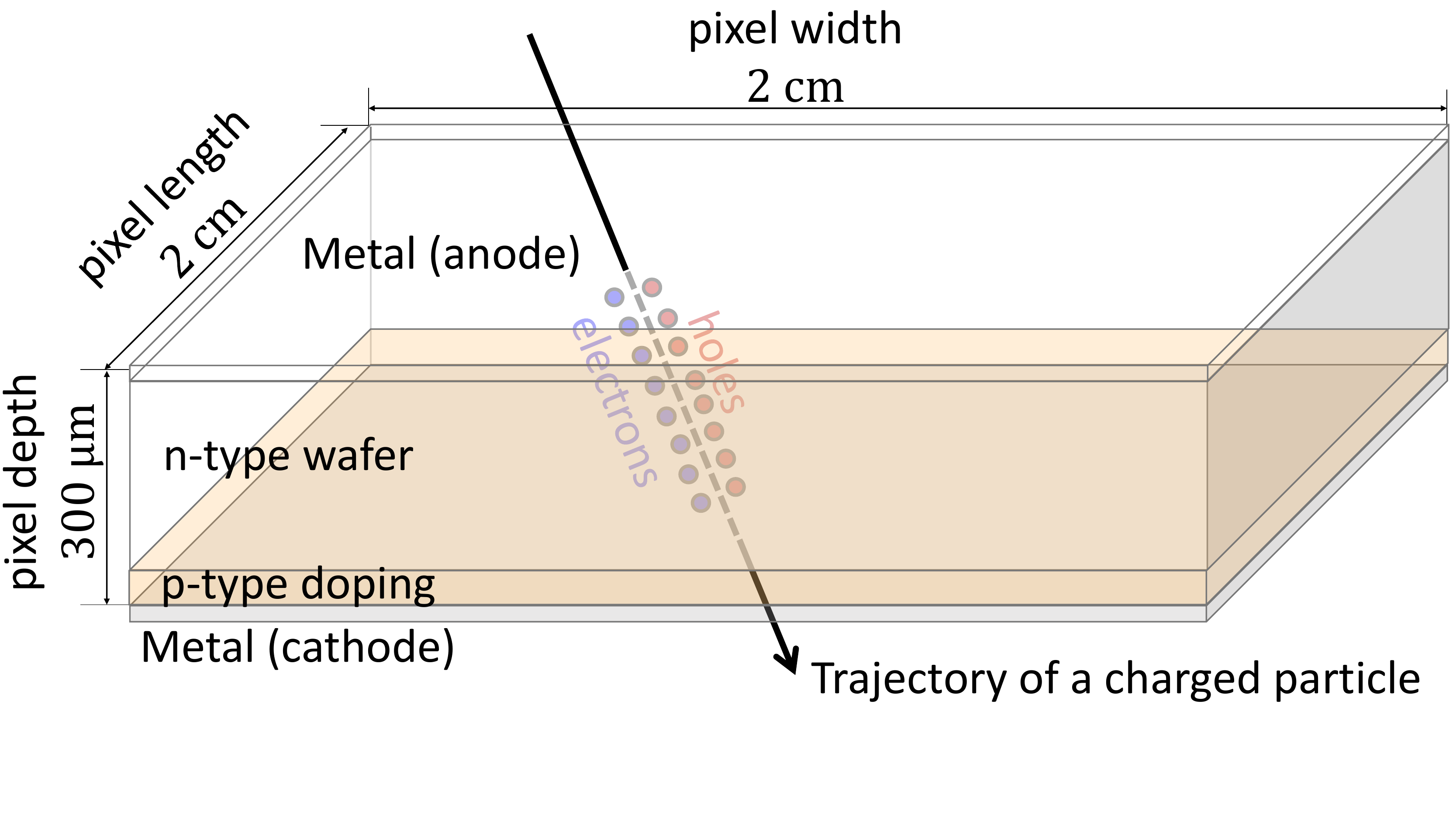}
\caption{(Color online) Schematic view of the silicon-pad detector in this study. The silicon-pad detector is \unit[2]{cm} wide and \unit[2]{cm} long. The depth of the detector is \unit[300]{$\mathrm{\mu}$m}. }
\label{fig:pixelschematic}
\end{figure}

The silicon detector has become one of the most important detectors recently in particle and nuclear physics because of its superior spatial and kinematic resolutions as well as the good response time than other types of detectors. The silicon semiconductor detector has the form of a p-n junction exerted with a reverse-bias voltage on each edge of the p- and n-type semiconductors. 

At the interface between the p-n junction, some of the electrons in the n-type semiconductor cross the junction and fill the holes in the p-type semiconductor and some of the holes in the p-type pass the junction and disappear with the n-type electrons. Therefore, the n-type region has a positive electric potential and the p-type region has a negative electric potential. This causes an electric field across the junction, forming a depletion layer without charge carriers near the interface. 

The silicon detector uses the depletion layer to obtain a current signal when charged particles pass through the volume. When a reverse-bias voltage is applied to the p-n junction, electrons and holes are forced toward the electrode, resulting in widening the depletion layer. A wider depletion layer increases the effective volume that is sensitive to detect charged particles. 

The LAMPS Collaboration~\cite{LAMPS} at RAON~\cite{RAON} designs a detector sensitive to energy measurement for low-energy heavy-ion collisions with enhanced isotopic identification capabilities. Charge and mass information of all reaction products is essential to advance the knowledge of such as the density dependence of nuclear symmetry energy, in-medium isovector transport properties of nuclear matter, and equation of state (EOS) modification due to cluster effects at very low densities. 

The schematic structure of the silicon-pad detector in this study is shown in Fig.~\ref{fig:pixelschematic}.
Each pad has a volume of silicon semiconductor with an area of \unit[$2\times2$]{cm$^2$} and a depth of \unit[300]{$\mu$m}. The n-type silicon semiconductor (wafer) is the base of the pad and p-type semiconductor is made by a doping process on the n-type wafer. 
The lightly doped n-type silicon bulk and the heavily doped p+ silicon pad form a p-n junction. In the bulk between the two electrodes, an electric field is formed by an external voltage. Since a reverse-bias voltage is applied to the detector, no current flows except for a small amount of leakage current. However, when charged particles pass through the detector, electrons in the valance band are excited to the conduction band. As a result, electron-hole pairs are created and those charge carriers move along the electric field to the electrode where a current is induced. The induced current can be measured to detect particles. The leakage current is dominated by thermally generated electron-hole pairs (thermal diffusion) and its magnitude is 10--100 times less than the induced current.  

The mechanism of electron-hole pair generation in the detector can be simplified by considering the minimum ionizing particles (MIP) as charged particles pass through the detector. MIP refers to charged particles having energy loss in the material. MIP can be observed when the kinetic energy of the particle is more than twice the invariant mass of the particle. Since the ionization loss of MIP depends only slightly on the momentum of the particles~\cite{segre1953experimental}, MIP produces free charge carriers evenly over the trajectory of the charged particles.

We have developed an improved simulation tool that is based on ROOT at CERN~\cite{ANTCHEVA20092499} and is implementing more functionalities for the existing simulation tool~\cite{KWON:2018csm}. The new tool called the ``Fast Silicon Device Simulation'' is optimized for simple structures of the silicon detector and allows faster and more accurate calculations than the previous tool. Fast Silicon Device Simulation is also designed to show the variation of physical observables and quantities dynamically when running the simulation. The detailed description for the program can be found here~\cite{fastsilicon}.

\section{Simulation}

\subsection{Electric potential calculation in Fast Silicon Device Simulation}
\label{sec:potential}

\begin{figure}[!htb]
\includegraphics[width=0.6\textwidth]{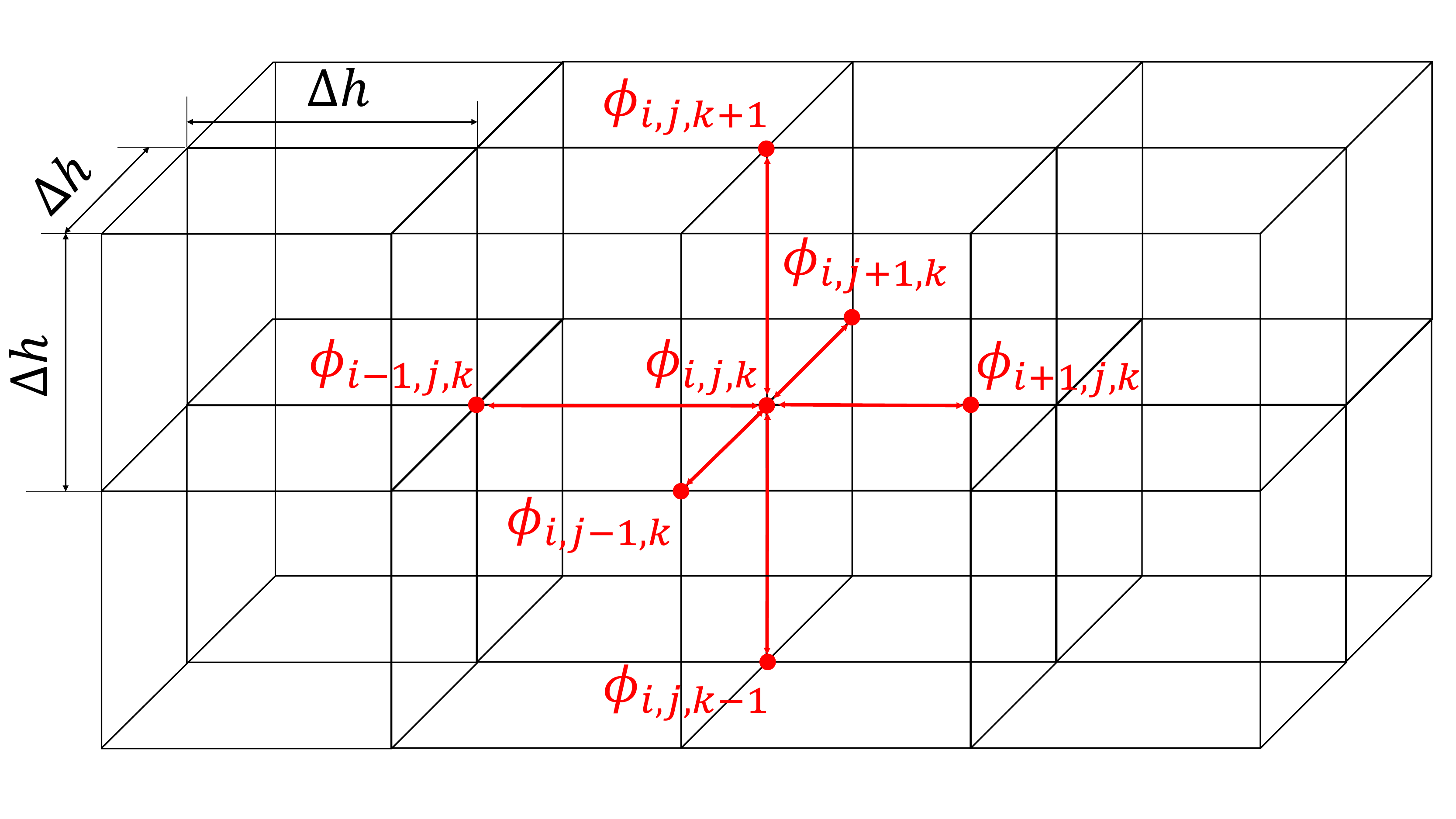}
\caption{(Color online) Virtual cubes for calculation of the potential are created. Each cube has a volume of $\Delta h^3$. The potential at the vertices is calculated consecutively and iteratively using the discrete Poisson equation in Eq.~\ref{eq:pebytaylor}.}
\label{fig:lattice}
\end{figure}

The Poisson equation is employed to calculate the electric potential in the detector. 
\begin{align}
    \hskip14em \nabla^2 \phi = - \rho / \epsilon \quad,
    \label{eq:pe}
\end{align}
where $\phi$ is the electric potential, $\rho$ is the volume charge density and $\epsilon$ is the permittivity of the material. Based on the boundary conditions of the detector, the potential calculated by Eq.~\ref{eq:pe} is called the drift potential.
As shown in Figure~\ref{fig:lattice}, three-dimensional lattice-type sub-structures of \unit[1]{$\rm \mu m$} in width, length and height virtually are made serially in the detector and electric potentials are calculated consecutively.

The left side of the Poisson equation can be expressed after expansion by Taylor series
\begin{align}
\begin{aligned}
     \hskip2em \nabla^2 \phi_{i,j,k} = \frac{1}{\Delta h^2} ( \phi_{i+1,j,k} &+ \phi_{i-1,j,k} + \phi_{i,j+1,k} \\& +  \phi_{i,j-1,k} + \phi_{i,j,k+1} + \phi_{i,j,k-1}- 6\phi_{i,j,k} )  
     + O(\Delta h^2) \quad,
    \label{eq:pebytaylor}
\end{aligned}
\end{align}
where $\phi_{i,j,k}$ is the potential at the vertex of the $i$, $j$ and $k^\mathrm{th}$ lattice along the directions of the width, length and depth of the device, respectively, and $\Delta h$ is the size of the lattice. Finally, $\phi_{i,j,k}$ is derived by substituting Eq.~\ref{eq:pebytaylor} to Eq.~\ref{eq:pe} as
\begin{align}
\begin{aligned}
    \hskip-1.5em \phi_{i,j,k} = \left( \phi_{i+1,j,k} + \phi_{i-1,j,k} + \phi_{i,j+1,k} + \phi_{i,j-1,k} + \phi_{i,j,k+1} + \phi_{i,j,k-1}\right)/6 + \Delta h^2 \left( \rho/\epsilon\right)/6 \quad.
    \label{eq:pebytaylorfinal}
\end{aligned}
\end{align}
The potential of the boundaries of the device is defined by the simulation user, and the internal potential inside the device is set to 0 initially. Calculating by substituting Eq.~\ref{eq:pebytaylor} for the entire lattice component, the component closest to the interface has a nonzero potential, which is closer to the expected potential. Once again, for the entire lattice component, the component closest to the interface has a value that is closer to the expected value than before, and has a nonzero potential for more internal components. If this process is repeated enough, the entire lattice component converges to a value that satisfies the above equation.

The iterative method of calculating the Poisson equation for all the components of the grid can be a good way to approximate the potential calculation, but it takes a very long time since many calculations are required to fill the inside from the grid boundary. Also, this method might fall into a local minimum that can be determined by special boundary conditions given by the nearby lattices. The time required for the iterative process and the local minimal problem can be significantly reduced by introducing the multi-grid method~\cite{KWON:2018csm, Weightingfield}. 

\begin{figure}[!htb]
\includegraphics[width=\textwidth]{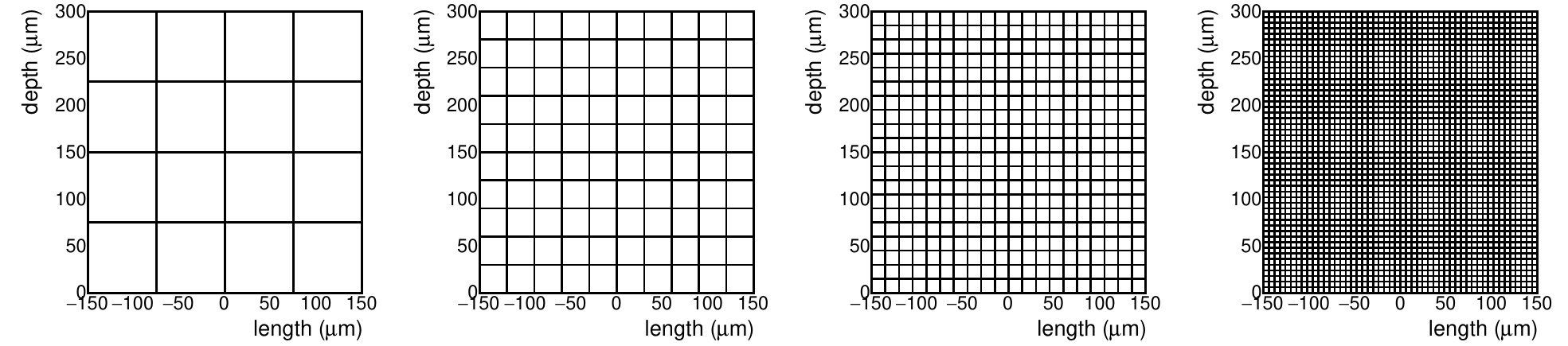}
\caption{Multi-grid method proceeds from a wide (left) to a narrow lattice (right) in order to improve the speed of the iteration method. For each step, the iteration method is repeated until the potential values at the lattices are stabilized. The multi-grid method with finer lattices applies the iteration method with interpolated potential values from the previous step at the beginning. Note that the figures show the full size along the direction of depth while a limited range (-150 to \unit[150]{$\mu$m}) is shown along the length direction of the device.} 
\label{fig:multigrid}
\end{figure}

The multi-grid method applies the iterative method for wider lattices as an initial condition. Then, denser lattices are created by interpolation with the average of the surrounding lattice components. At each finer lattice process, the iteration method is repeated until values of potential are stabilized and the size of the lattices becomes \unit[1]{$\mu m$}. Unlike the general iteration method, which gradually fills the components from the grid boundary to the inside, the multi-grid method fills the inside of the grid with the expected value (average value) and then proceeds with the iteration method, so the number of calculations to satisfy Eq.~\ref{eq:pe} is considerably decreased~\cite{KWON:2018csm, Weightingfield}. Fig.~\ref{fig:multigrid} shows how the iterative and multi-grid methods are applied.

\subsection{Electric field calculation in Fast Silicon Device Simulation}

The electric field is calculated by applying the equation $\vec{E} = -\nabla \phi$ to the result of the potential obtained by the method described in Sec~\ref{sec:potential}. The following calculation is made for all array components of the lattices to obtain the electric field
\begin{align}
    \label{eq:Efield}
    \begin{aligned}
    \hskip12em E_x &= - \frac{\phi_{i,j,k}-\phi_{i-1,j,k}}{\Delta h} \\
    E_y &= - \frac{\phi_{i,j,k}-\phi_{i,j-1,k}}{\Delta h} \\
    E_z &= - \frac{\phi_{i,j,k}-\phi_{i,j,k-1}}{\Delta h} \quad,
    \end{aligned}
\end{align}
where $E_x$, $E_y$ and $E_z$ are the electric field components in the directions of the width, height and the length of the device, respectively. 

\subsection{Induced current calculation in Fast Silicon Device Simulation}
The induced current calculation in the simulation is based on so-called the drift-diffusion model~\cite{Leo:1987kd}, which can be described by 
\begin{align}
\label{eq:df}
    \begin{aligned}
   \hskip12em \vec{J}_\mathrm{e} &= e \rho_\mathrm{e} \mu_\mathrm{e} \vec{E} + e D_\mathrm{e} \nabla n_\mathrm{e} \\
    \vec{J}_\mathrm{h} &= e \rho_\mathrm{h} \mu_\mathrm{h} \vec{E} - e D_\mathrm{h} \nabla n_\mathrm{h} \quad,
    \end{aligned}
\end{align}
where $\vec{J}_\mathrm{e}$ ($\vec{J}_\mathrm{h}$) is the current density of electrons (holes), $\rho_\mathrm{e}$ ($\rho_\mathrm{h}$) is the volume charge density of electrons (holes), $e$ is the elementary charge, $\mu_\mathrm{e}$ ($\mu_\mathrm{h}$) is the mobility of electrons (holes) in the device, $n_\mathrm{e}$ ($n_\mathrm{h}$) is the electron (hole) density, and $D_\mathrm{e}$ ($D_\mathrm{h}$) is the Einstein's relation that can be expressed as $D_\mathrm{e} = - k_\mathrm{B}T\mu_\mathrm{e}/e$ ($D_\mathrm{h} = k_\mathrm{B}T\mu_\mathrm{h}/e$) for electrons (holes), $k_\mathrm{B}$ is the Boltzmann constant, and $T$ is the operational temperature of the device. $\rho_{\rm e}$ ($\rho_{\rm h}$) is the volume charge density for electrons (holes).

The calculation of the induced current in the simulation analyzes the Poisson equation in Eq.~\ref{eq:pe}, drift-diffusion equation in Eq.~\ref{eq:df}, and current continuity equation in Eq.~\ref{eq:con} simultaneously. The current continuity equation describes the continuity of current in the closed curved area surrounding a specific volume following the law of charge conservation
\begin{align}
\label{eq:con}
    \begin{aligned}
   \hskip11em \frac{\partial \rho_\mathrm{e}}{\partial t} - \frac{1}{e} \vec{\nabla}\cdot \vec{J_\mathrm{e}} &= G_\mathrm{e} - R_\mathrm{e}  \\
    \frac{\partial \rho_\mathrm{h}}{\partial t} + \frac{1}{e} \vec{\nabla}\cdot \vec{J_\mathrm{h}} &= G_\mathrm{h} - R_\mathrm{h}  \quad,
    \end{aligned}
\end{align}
where $G_\mathrm{e}$ ($G_\mathrm{h}$) is the generation rate of electrons (holes), and $R_\mathrm{e}$ ($R_\mathrm{h}$) is the recombination rate of electrons (holes). In Fast Silicon Device Simulation, the recombination rate is ignored as the effect is found to be very small. The generation rate is given only at $t=0$ when the simulation starts. The generation rate is fixed by the MIP, incident angle and position of a charged particle passing through the device at $t=0$.  

In Fast Silicon Device Simulation, the mobility of charge carriers can be expressed in terms of the electric field. This study follows Caughey and Thomas Expression~\cite{CaugheyThomas} 
\begin{align}
\label{eq:mob}
    \hskip3em \mu_\mathrm{e}(E) = \mu_\mathrm{0,e} \left( \frac {1} { 1+\left(\frac {\mu_\mathrm{0,e} E}{v_{\rm sat,e} }  \right)^2} \right) ^ {1/2}  \mathrm{and} \quad \mu_\mathrm{h}(E) = \mu_\mathrm{0,h} \frac {1} { 1+\frac {\mu_\mathrm{0,h} E}{v_{\rm sat,h} }  } \quad,
\end{align}
 where $\mu_\mathrm{e}(E)$ ($\mu_\mathrm{h}(E)$) is an electric-field dependent mobility for electrons (holes) with a constant $\mu_\mathrm{0,e} = \unit[1350]{\mathrm{cm}^2/Vs}$ ($\mu_\mathrm{0,h} = \unit[480]{\mathrm{cm}^2/Vs}$), and where $v_{\rm sat, e}$ ($v_{\rm sat, h}$) is the saturation velocity of electrons (holes). The saturation velocity of electrons (holes) is fixed as $v_{\rm sat, e}= 1.1\times 10^{5}\,\mathrm{m/s}$ ( $v_{\rm sat, h}= 9.5\times 10^{4}\,\mathrm{m/s}$) at \unit[$T = 300$]{K}.


\subsection{Silvaco TCAD and Garfield++ simulation}
Induced current results simulated by Fast Silicon Device Simulation are compared to those by Silvaco~\cite{SilvacoAtlas} TCAD and Garfield++~\cite{Garfieldpp} simulations.
Semiconductor device operation can be simulated by Technology Computer-Aided Design (TCAD).
Silvaco TCAD includes a text editor ($Devedit$) and a runtime engine ($Deckbuild$) which writes a structure file. The $Atlas$ is a device simulator for optimization and characterization of semiconductor devices. The mobility models, semiconductor physical models, biasing conditions, etc. are implemented through $Atlas$ and the configuration of a simulation is controlled with $Devedit$ by inserting parameters and models. A model following Eq.~\ref{eq:mob} is selected. The signal is calculated using Eq.~\ref{eq:con} in $Atlas$ device simulation.

Garfield++ is a simulation toolkit for particle detectors that utilize gas or semi-conductors for sensitive medium. 
In Garfield++, the current signals are calculated using the Shockley-Ramo theorem~\cite{Shockley,Ramo}. The induced current by charge carriers at a specific time $t$ is given by
\begin{align}
    \hskip7em I(t) = - \int q(x, y, z, t)\vec{v}(t)\cdot \vec{E}_{w}(x, y, z, t)  \mathrm{d}x\mathrm{d}y\mathrm{d}z \quad,
\end{align}
where $q$ and $\vec{v}$ are the total charge and average velocity vector of charge carriers at the position of $(x, y, z)$ in the device volume, respectively, and $\vec{E}_{w}$ is the weight field. The weighting field $\vec{E}_{w}$ is calculated using Eq.~\ref{eq:Efield} with conditions that the potential at the anode is the unit potential and the potential at the cathode is zero~\cite{Riegler_2016}. 
For the Garfield++ simulation in this study, the electric field and the weighting field $\vec{E}_{w}$ are obtained from Silvaco TCAD. The mobility is modified to follow Eq.~\ref{eq:mob} with the same values of $\mu_\mathrm{0,e}$, $\mu_\mathrm{0,h}$, $v_{\rm sat, e}$ and $v_{\rm sat, h}$ for the Garfield++ simulation in this study.

\section{Simulation setup in Fast Silicon Device Simulation}

\begin{table}[!htb]
\footnotesize
\begin{tabular}{ c|c|c|c } 
\hline

\hline
    & \quad Fast Silicon Device Simulation \quad & \quad  TCAD \quad  &  \quad Garfield++ \quad \\
\hline

\hline
Device dimension (width$\times$length$\times$depth) & \multicolumn{3}{c}{\unit[2]{cm}$\times$\unit[2]{cm}$\times$\unit[300]{$\mu$m} }\\
\hline 
Reverse-bias voltage &  \multicolumn{3}{c}{\unit[140]{V}}  \\
\hline 
Doping density (n-type) &  \multicolumn{2}{c|}{\unit[$1.9\times10^{18}$]{m$^{-3}$}} & -- \\
\hline
Doping density (p-type)  &  \multicolumn{2}{c|}{\unit[$10^{21}$]{m$^{-3}$}} & -- \\
\hline
Operational temperature ($T$) &   \multicolumn{3}{c}{\unit[300]{K} } \\
\hline
Vacuum permittivity ($\epsilon_0$) & \multicolumn{3}{c}{\unit[$8.854\times 10^{-12}$]{$\rm F\,m^{-1}$}} \\
\hline
Relative permittivity of Si ($\epsilon_\mathrm{r}$) & \multicolumn{3}{c}{11.68} \\
\hline
Mobility constant for electrons ($\mu_\mathrm{0,e}$) & \multicolumn{3}{c}{\unit[0.135]{m$^2\,$V$^{-1}$s$^{-1}$}} \\
\hline
Mobility constant for holes ($\mu_\mathrm{0,h}$) &  
\multicolumn{3}{c}{\unit[0.048]{m$^2\,$V$^{-1}$s$^{-1}$}} \\
\hline
Saturation velocity of electrons ($v_{\mathrm{sat,e}}$) & \multicolumn{3}{c}{\unit[$1.1\times10^5$]{m$\,s^{-1}$}} \\
\hline
Saturation velocity of holes ($v_\mathrm{sat,h}$) & \multicolumn{3}{c}{\unit[$9.5\times10^4$]{m$\,$s$^{-1}$}} \\
\hline
Mobility of electrons ($\mu_\mathrm{e}$) & \multicolumn{3}{c}{$ \mu_\mathrm{0,e} \left( \frac {1} { 1+\left(\frac {\mu_\mathrm{0,e} E}{v_{\rm sat,e} }  \right)^2} \right) ^ {1/2}$ } \\
\hline
Mobility of holes ($\mu_\mathrm{e}$) & \multicolumn{3}{c}{$ \mu_\mathrm{0,h} \frac {1} { 1+\frac {\mu_\mathrm{0,h} E}{v_{\rm sat,h} }  } $ }\\
\hline
MIP $\left( \int G_\mathrm{e,h} \mathrm{d}t / \Delta r \right) $ &  \multicolumn{3}{c}{\unit[75]{$\rm \mu m^{-1} $}} \\
\hline
\end{tabular}
\caption{Input parameters and models for Fast Silicon Device Simulation, TCAD and Garfield++. Note that the absolute permittivity ($\epsilon$) of silicon in Eq.~\ref{eq:pe} is given by $\epsilon = \epsilon_{0} \epsilon_\mathrm{r}$. The linear density of electron-hole pairs generated by an incident particle (MIP) is given to Eq.~\ref{eq:con} as $\int G_\mathrm{e,h} \mathrm{d}t / \Delta r = \mathrm{MIP}$, where $r$ is the direction that the charged particle moves in the device.} 
\label{tab:inputparam}
\end{table}

\begin{figure}[!htb]
\includegraphics[width=0.9\textwidth]{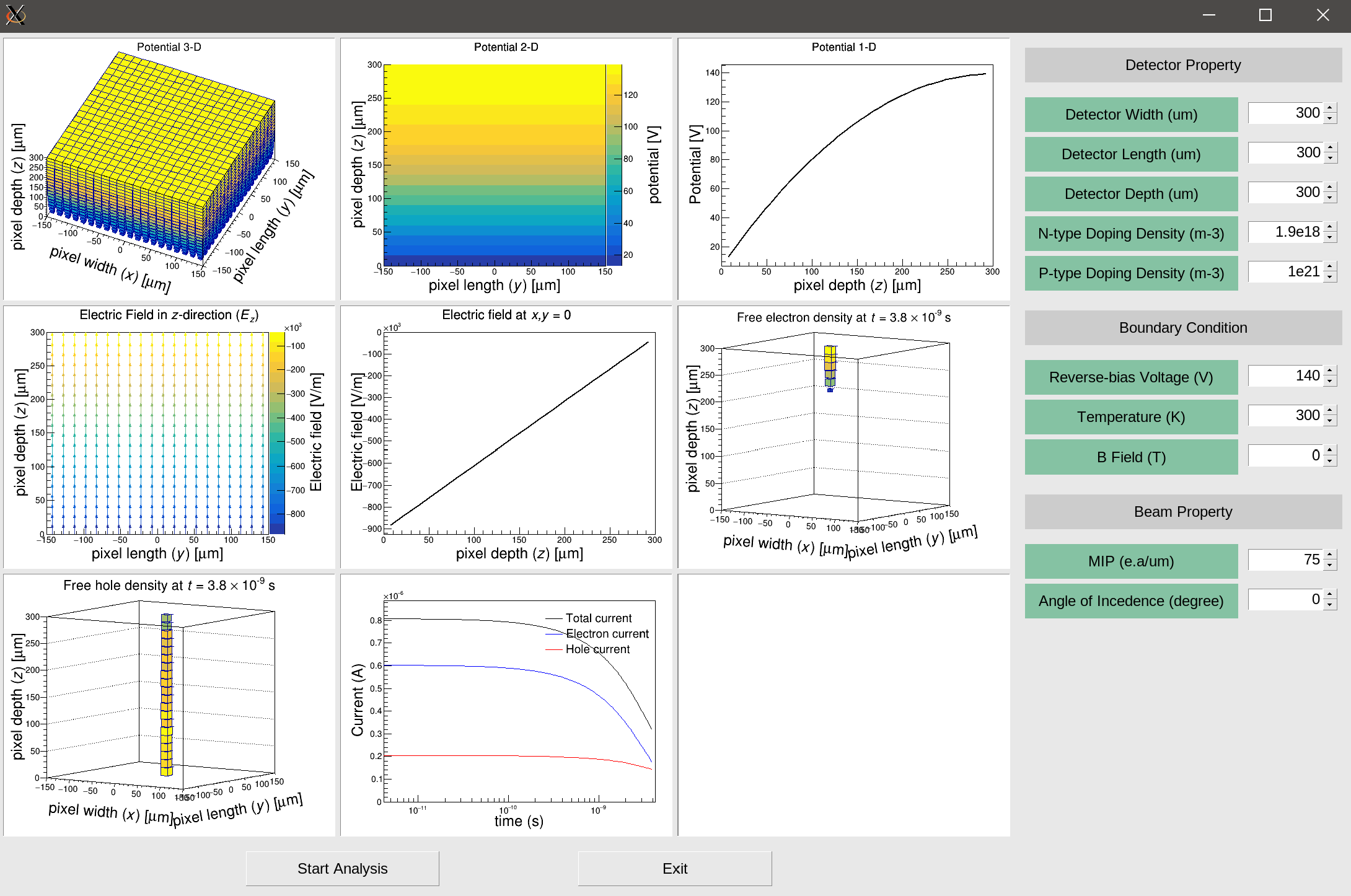}
\caption{(Color online) The application panel of Fast Silicon Device Simulation. Fast Silicon Device Simulation obtains values of detector properties, boundary conditions and beam properties from the interactive input boxes in the panel. The simulation starts when the button ``Start Analysis'' is clicked.   }
\label{fig:fastsiliconpanel}
\end{figure}

Table~\ref{tab:inputparam} shows all input parameters and used models for Fast Silicon Device Simulation, TCAD and Garfield++. 
The Fast Silicon Device Simulation calculates the potential, electric field and induced current with the input values controlled by the GUI as shown in Fig~\ref{fig:fastsiliconpanel}. The input values are the dimension of the device, the reverse-bias voltage exerted on the device, the operating temperature of the device, and the angle of incidence and generation rate of electron-hole pairs for a charged particle passing through the device. The reverse-bias voltage of \unit[140]{V} is determined so that the length ($W$) of the depletion region is same as the depth of the device by employing the equation
\begin{equation}
   \hskip15em W = \sqrt{\frac{2 \epsilon V}{e N}} \quad,
\end{equation}
where $V$ is the reverse-bias voltage exerted on the device, $e$ is the elementary charge and $N$ is the doping density of the bulk wafer~\cite{Leo:1987kd}.

The dimension specifies the width, length and depth of the detector. The reverse-bias voltage designates the potential difference between the the two electrodes with a larger potential value at the anode than the cathode. The input value of the incident angle fixes the angle between the directions of the incident particle and the depth. Electron-hole pairs are uniformly generated along the path of an input particle. The linear density of the integrated generation rates of electron-hole pairs is expressed as $\int G_\mathrm{e,h} \mathrm{d}t / \Delta r$, where $r$ is the direction of the path and its value at $t=0$ is set by $\rm 75\,\mu m^{-1}$ for simulation in this study. The operating temperature of the detector can be input by entering the value. The default value of the temperature is \unit[300]{K}. The detailed information of Fast Silicon Device Simulation can be found in the developing website~\cite{fastsilicon}.

\section{Results}

\begin{figure}[hbt!]
\subfloat[]{%
   \includegraphics[width=0.48\linewidth]{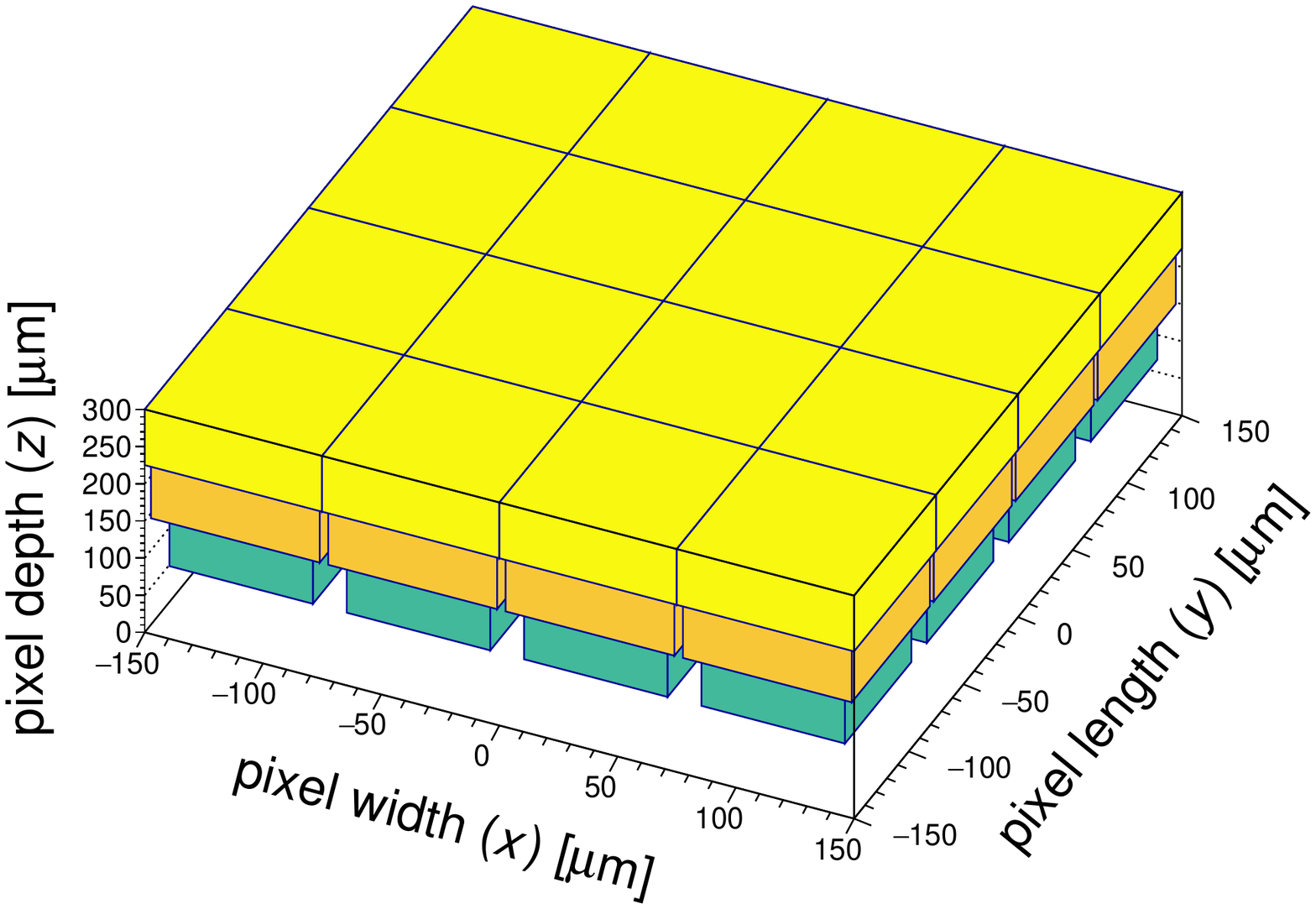}
}
\subfloat[]{%
    \includegraphics[width=0.48\linewidth]{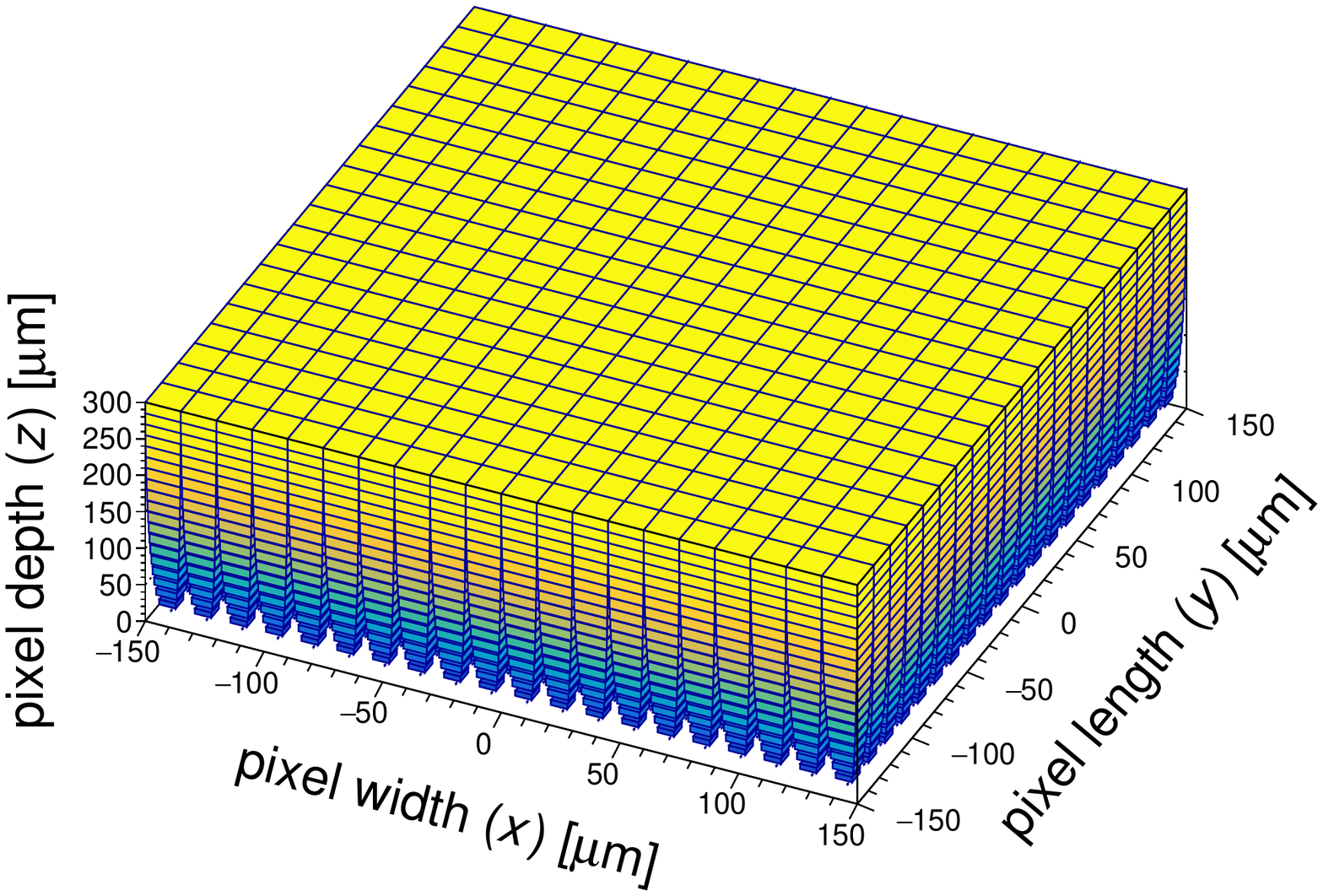}
}

   \caption{ The potential calculation of the multi-grid method is applied sequentially from wider (a) to narrower size (b) of the lattice to solve the Poisson equation quickly. The color and the size of the box in each bin show the magnitude of the calculated potential. Note that the range of the directions in pad width ($x$) and length ($y$) is only shown for the part from -150 to \unit[150]{$\mathrm{\mu m}$} instead of the actual size from -10000 to \unit[10000]{$\mathrm{\mu m}$}.} 
  \label{fig:resultpotential2}
\end{figure}

\begin{figure}[hbt!]
\subfloat[]{%
    \includegraphics[width=0.48\linewidth]{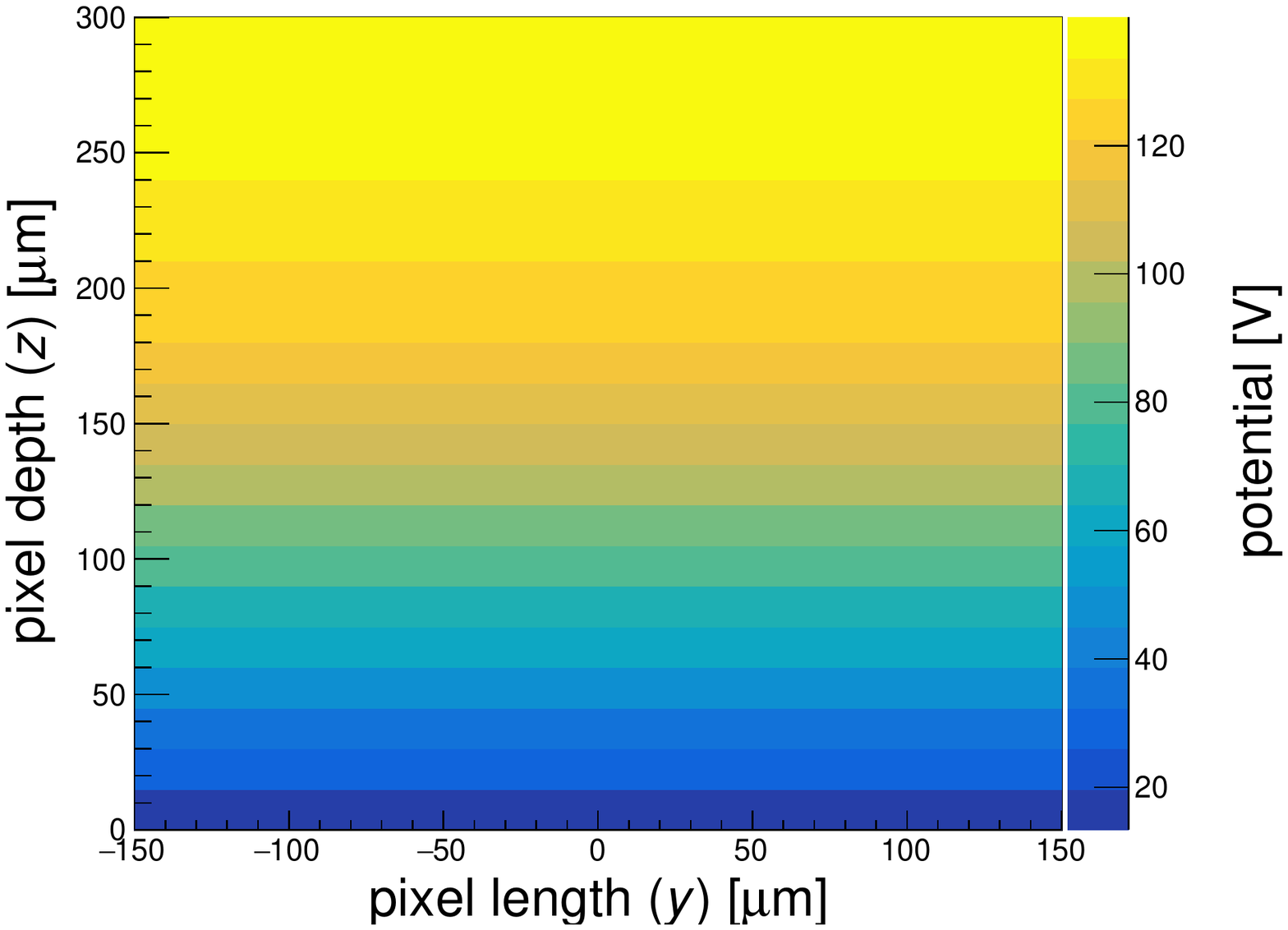}
}
\subfloat[]{%
	\includegraphics[width=0.48\linewidth]{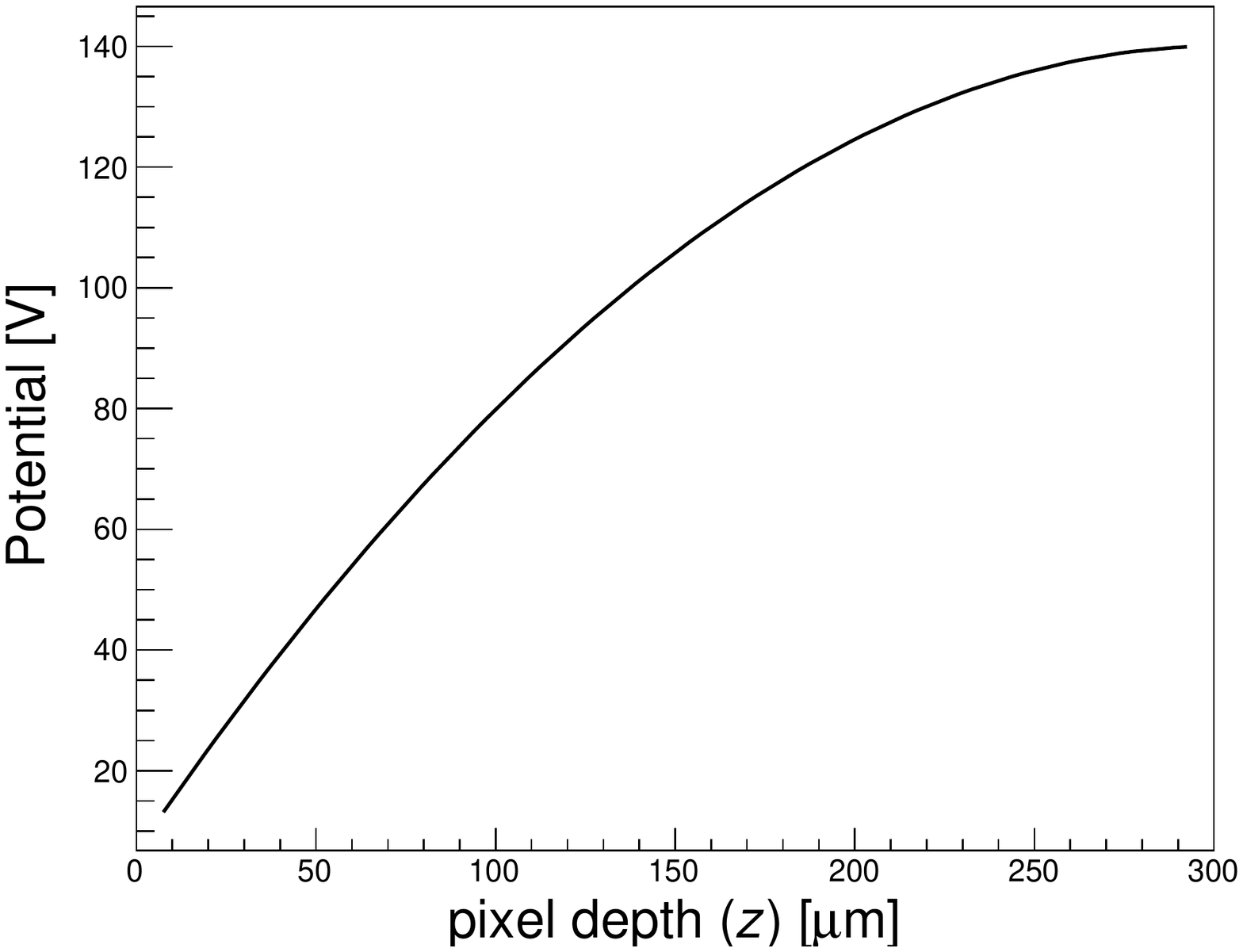}
}

  \caption{The left panel shows the cross-sectional view of the potential values on the $yz$ plane at $x=0$ for Fig.~\ref{fig:resultpotential2} (b). The right panel shows values along the $z$-axis at $x,y=0$ of the left panel when a reverse bias voltage of \unit[140]{V} is exerted on the device. }
  \label{fig:resultpotential3}
\end{figure}

The potential calculation with the multi-grid method is shown in Figure~\ref{fig:resultpotential2}. The method proceeds from a wider (Fig.~\ref{fig:resultpotential2}a) to the narrower lattice (Fig.~\ref{fig:resultpotential2}b) in order to improve the speed of the iteration method. For each step, the iteration method is repeated until the potential values at the lattices do not change by less than 0.01\%. When the multi-grid method moves to a finer lattice, the iteration method starts at the beginning with the interpolated potential values from the previous wider lattice. The color and box size of Fig.~\ref{fig:resultpotential2} represent the magnitude of the calculated potential values. Also, it is worth noting that the figures show the full size along the direction of depth while a limited range (-150 to \unit[150]{$\mu$m}) is only shown along with the length and width directions of the device instead of the full size ranging from -10000 to \unit[10000]{$\mu$m}.

The calculated electric potential with a reverse-bias voltage of \unit[140]{V} is shown in Fig.~\ref{fig:resultpotential3}. The left panel represents the potential values projected on $yz$-plane at $x=0$, where the $z$-axis is the direction in the depth of the device in Fig.~\ref{fig:lattice}. The right panel shows the projected values of the left panel at $x,y = 0$. The potential distribution along the direction of depth in a depleted p-n junction volume is expected to be a second-order polynomial and the trend is simulated well by Fast Silicon Device Simulation as expected.

\begin{figure}[hbt!]
\subfloat[]{%
    \includegraphics[width=0.48\linewidth]{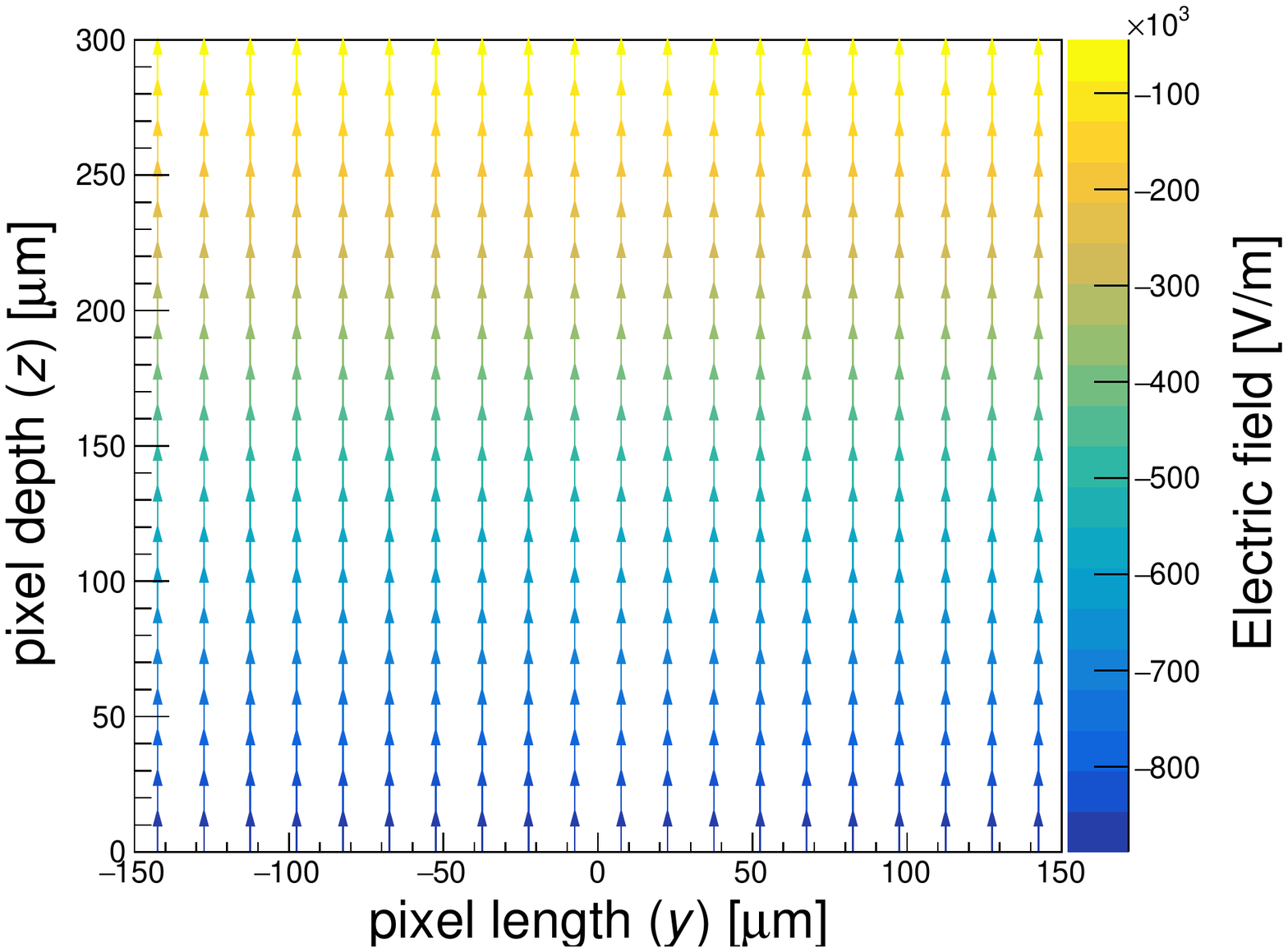}
}
\subfloat[]{%
	\includegraphics[width=0.48\linewidth]{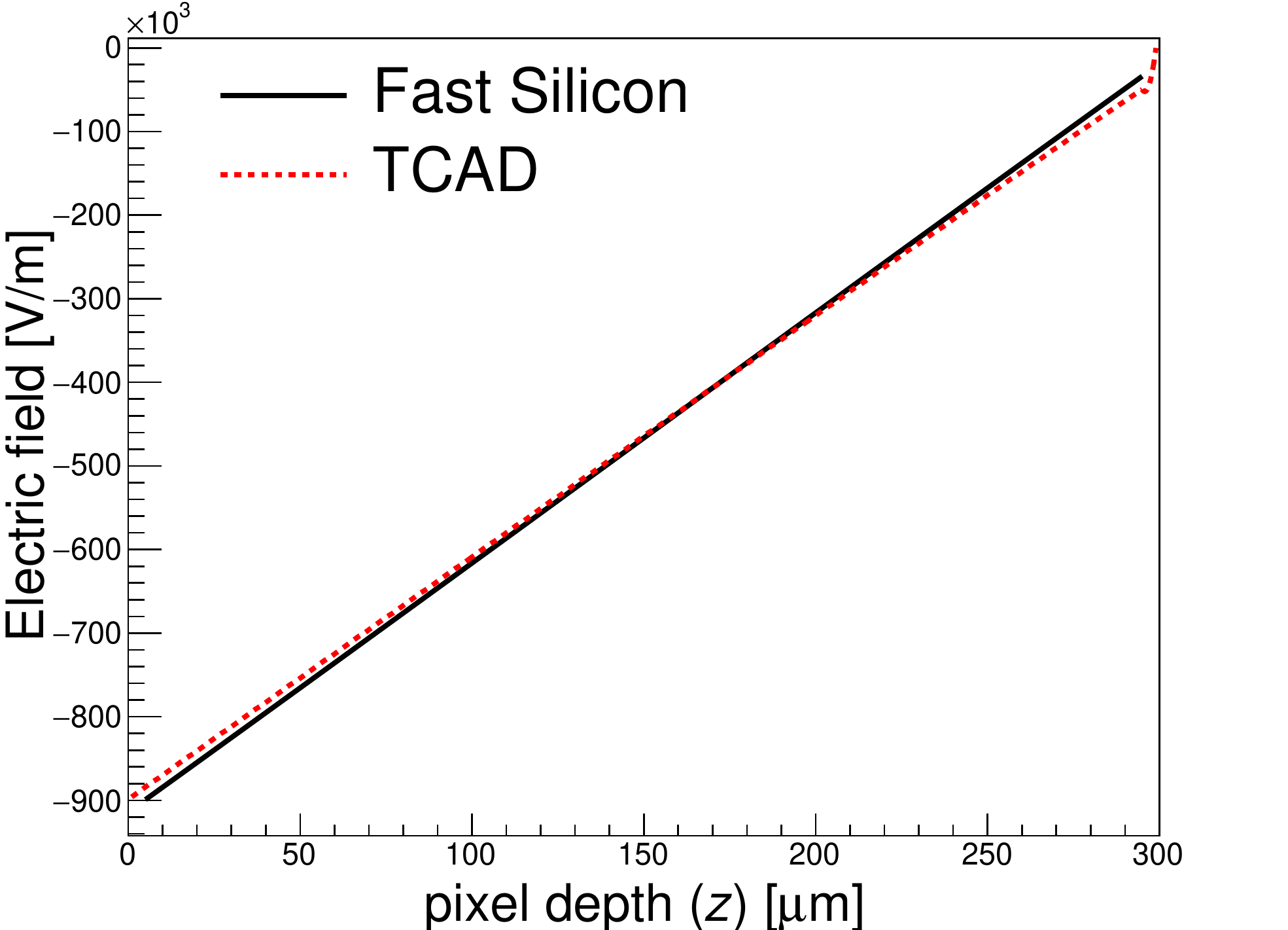}
}

  \caption{(a) The direction and magnitude of the electric field on $yz$-plane at $x=0$. (b) The electric field calculated in Fast Silicon Device Simulation (Fast Silicon) along the $z$-axis at $x,y=0$. The result is compared with TCAD calculation.}
  \label{fig:efield}
\end{figure}

The electric field simulated by Fast Silicon Device Simulation is presented in Fig.~\ref{fig:efield}. The left panel shows the direction and magnitude of the electric field on $yz$-plane at $x=0$. 
The right panel in Fig.~\ref{fig:efield} shows the projected electric field along the $z$-axis at $x,y=0$ of the left figure. The device is reverse-biased over its entire depth of $300\,\mu\mathrm{m}$. 
As a result, the electric field is expected to vary linearly from 0 to $300\,\mu\mathrm{m}$ because of the constant doping density (\unit[$1.9\times10^{18}$]{$\mathrm{m^{-3}}$}) in the bulk n-type silicon.  
Fast Silicon Device Simulation simulates electric field as expected and the result is ensured by 
the result with Silvaco TCAD as shown in Fig.~\ref{fig:efield}.

\begin{figure}[t]
\includegraphics[width=0.6\textwidth]{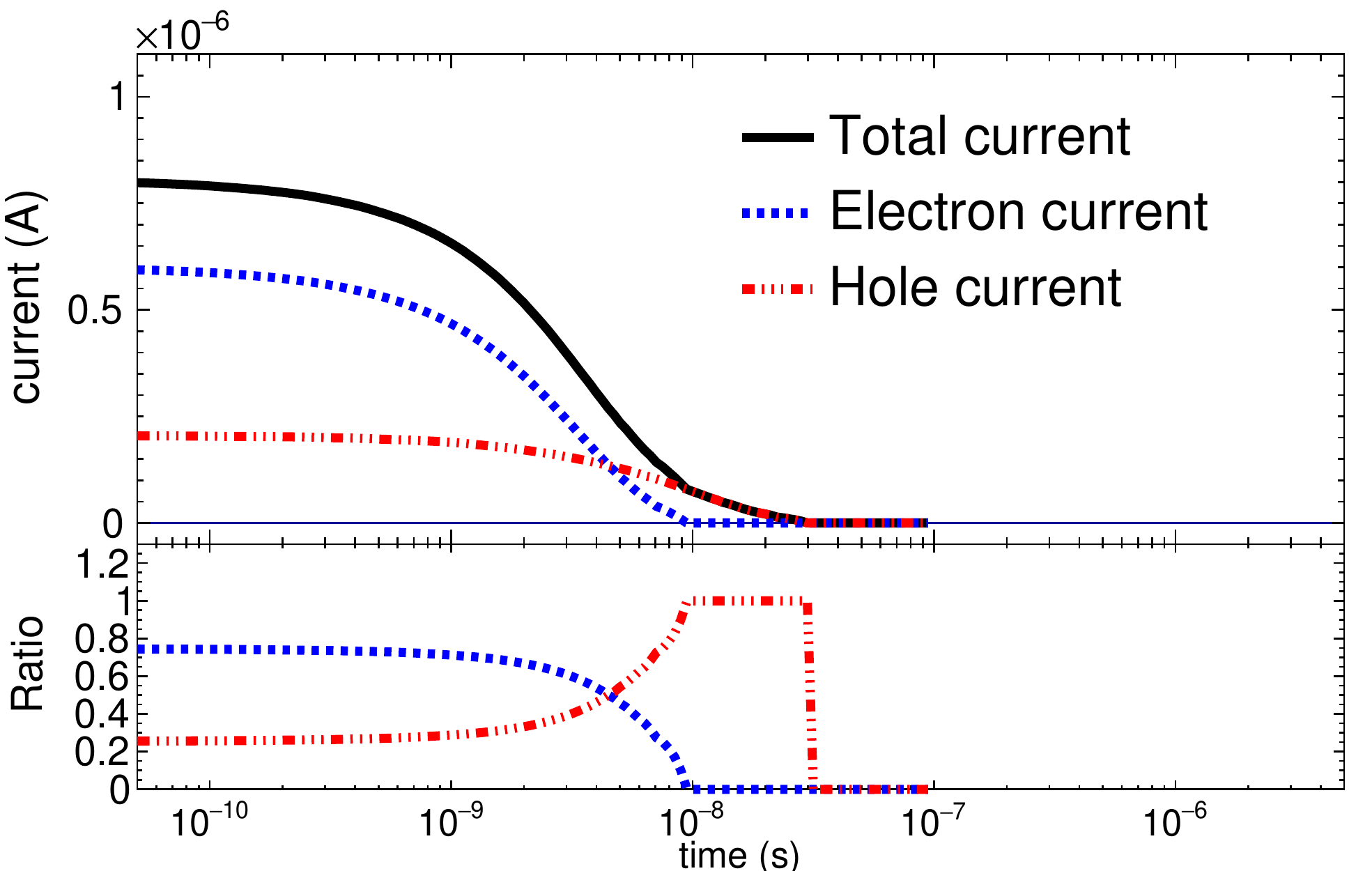}
\caption{(Color online) Induced currents in Fast Silicon Device Simulation for generated electron-hole pairs along the passage of an incident particle. The total current is the sum of electron current and hole current. The bottom panel shows relative rates of electron and hole currents to the total current.}
\label{fig:curcomparison}
\end{figure}

Transient currents induced by electrons and holes in Fast Silicon Device Simulation as a function of time are shown in Fig.~\ref{fig:curcomparison}. The electron and holes are generated along the trajectory of an incident charged particle. The number of generated electron-hole pairs is set by 75 per $\mu\mathrm{m}$ in the silicon volume as shown in Tab~\ref{tab:inputparam}. The particle trajectory is set from the top ($x$, $y$, $z$)=(0,0,0) to bottom (0,0,300$\mu\mathrm{m}$). The total current is the sum of the electron and hole currents. Assuming from the mobility constants of electrons and holes in Tab~\ref{tab:inputparam}, holes move slower than electrons in a silicon device. This effect makes the magnitude of the hole current is finally lower than that of the electron current. After all of the electrons are quickly collected to the anode, the hole current begins to account for most of the total current because of the same reason. 

\begin{figure}[!htb]
\includegraphics[width=0.6\textwidth]{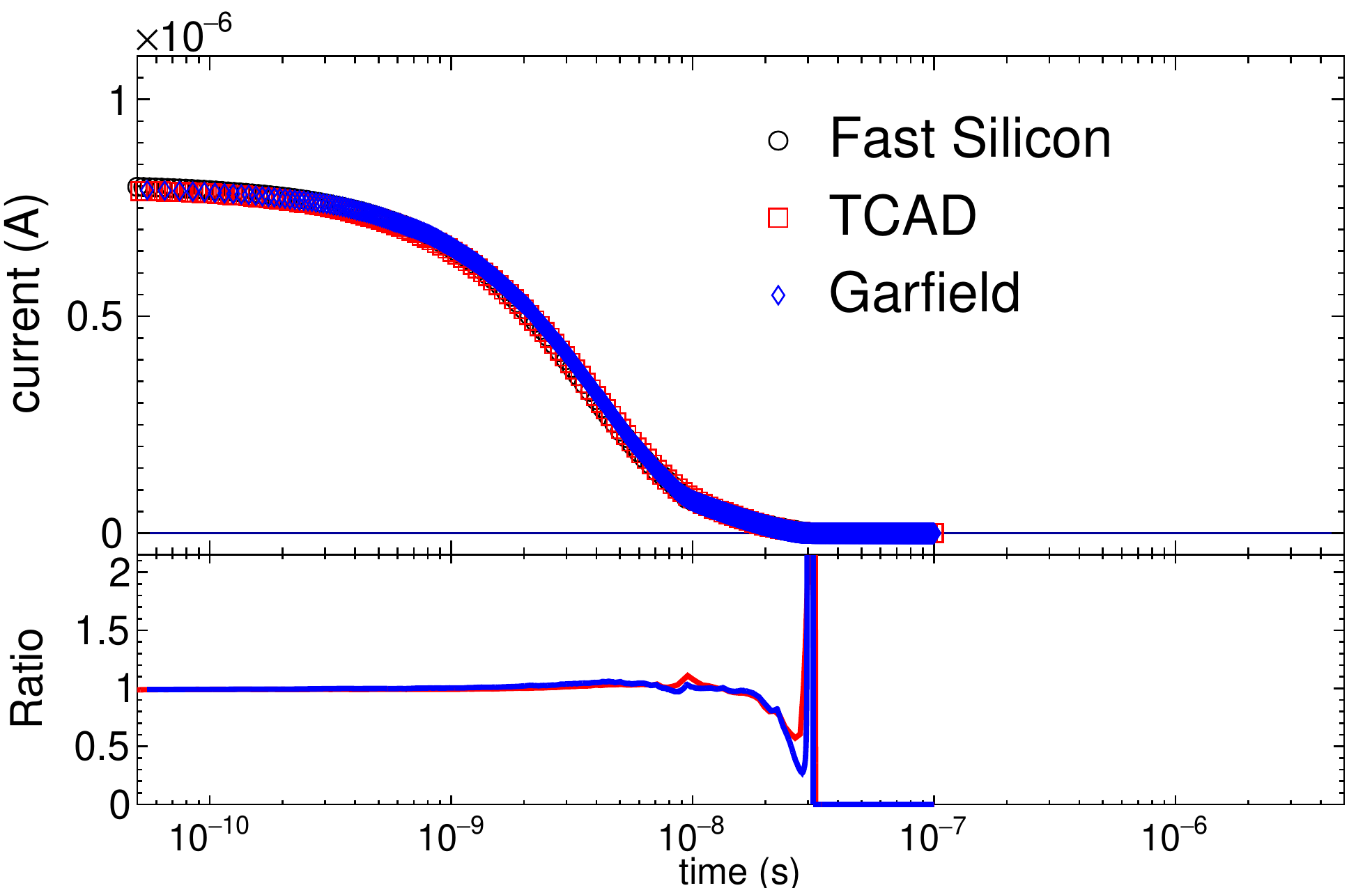}
\caption{(Color online) Total currents in Fast Silicon Device Simulation (Fast Silicon), TCAD and Garfield++. The bottom panel shows ratios of the currents in the upper panel. The red (blue) solid line in the ratio plot indicates $I_{\rm TCAD}/I_{\rm Fast\,Silicon}$ ($I_{\rm Garfield++}/I_{\rm Fast\,Silicon}$).}
\label{fig:curcomparison2}
\end{figure}

The total current simulated by Fast Silicon Device Simulation is compared with those of Silvaco TCAD and Garfield++ in the upper panel of Fig.~\ref{fig:curcomparison2}. The current signal by Fast Silicon Device Simulation is almost the same as those of TCAD and Garfield++ as shown in the ratio plot. The bottom panel shows ratios of the currents in the upper panel. The red (blue) solid line in the ratio plot represents $I_{\rm TCAD}/I_{\rm Fast\,Silicon}$ ($I_{\rm Garfield++}/I_{\rm Fast\,Silicon}$) where $I_{\rm Fast\,Silicon}$, $I_{\rm TCAD}$ and $I_{\rm Garfield++}$ are the total currents as a function of time for Fast Silicon Device Simulation, TCAD and Garfield++, respectively. The result from Fast Silicon Device Simulation is almost similar to the results from TCAD and Garfield++ when the same mobility model in Eq.~\ref{eq:mob} is used. This supports that results by Fast Silicon Device Simulation is reliable and all methods implemented in Fast Silicon Device Simulation is accurate in detail.

\begin{figure}[!htb]
\includegraphics[width=0.6\textwidth]{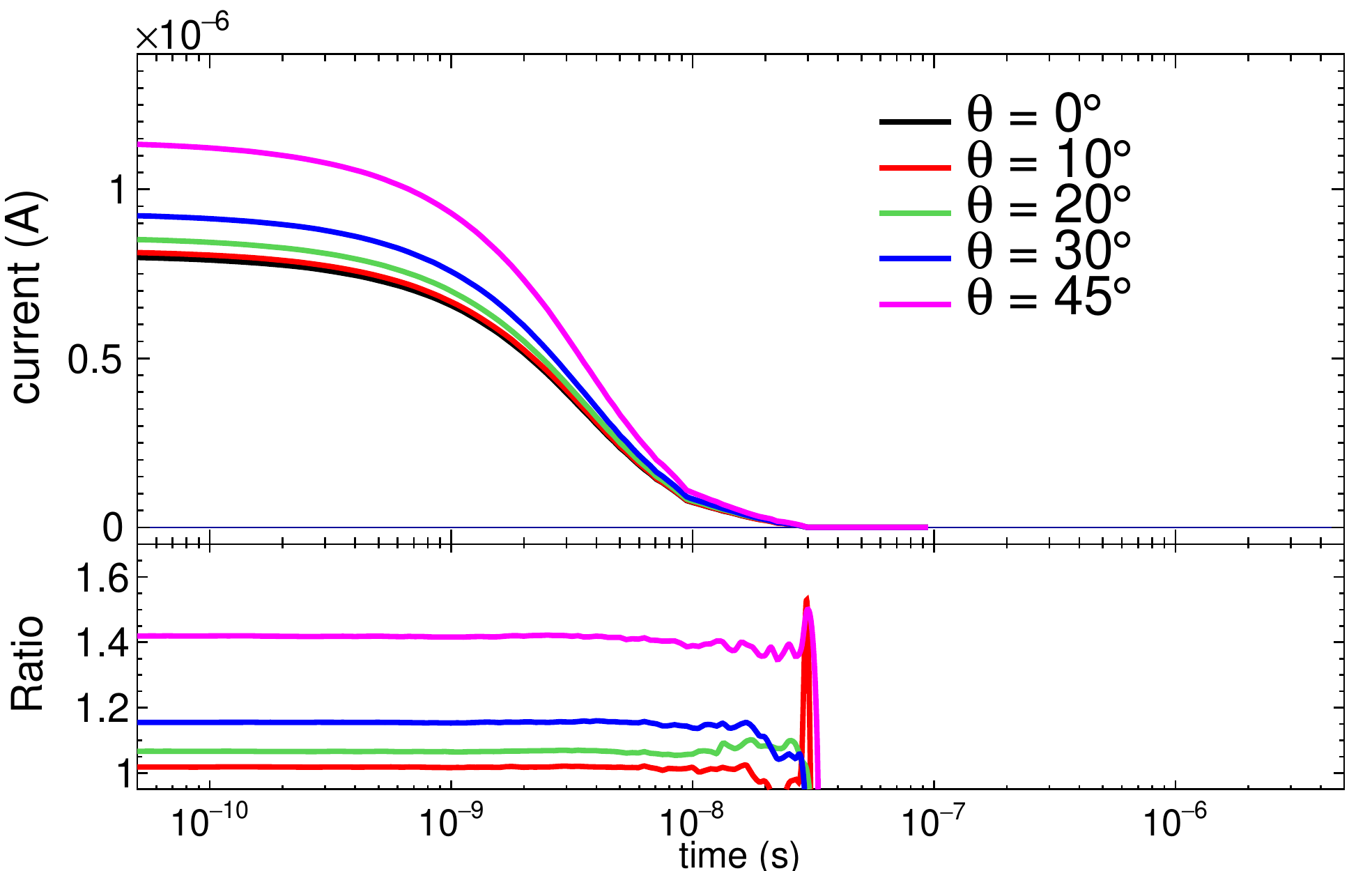}
\caption{(Color online) Total currents by varying the angle of incidence ($\theta$) from 0 to $45^{\circ}$ in Fast Silicon Device Simulation. The bottom panel indicates the ratio of the current for each incidence angle with respect to the case when $\theta = 0^{\circ}$.}
\label{fig:curcomparison3}
\end{figure}

Fast Silicon Device Simulation provides also a functionality to simulate the case when the incidence angle ($\theta$) of a charged particle passing through the device varies. Fig.~\ref{fig:curcomparison3} shows incidence angle dependence of the current signal simulated by Fast Silicon Device Simulation. As the angle varies, the distance that a charged particle travels in the silicon volume increases by 1/$\cos{\theta}$. The current signal is expected to increase linearly for the increased distance because the number of generated electron-hole pairs is set to be constant (75 per $\mu\mathrm{m}$) in the silicon volume for the simulation. The maximum values of the currents for different incidence angles in the figure follow the 1/$\cos{\theta}$ dependence as expected. Also, it is confirmed that the incidence angle of a charged particle does not change the shape of the current. 

\begin{figure}[!htb]
\includegraphics[width=0.6\textwidth]{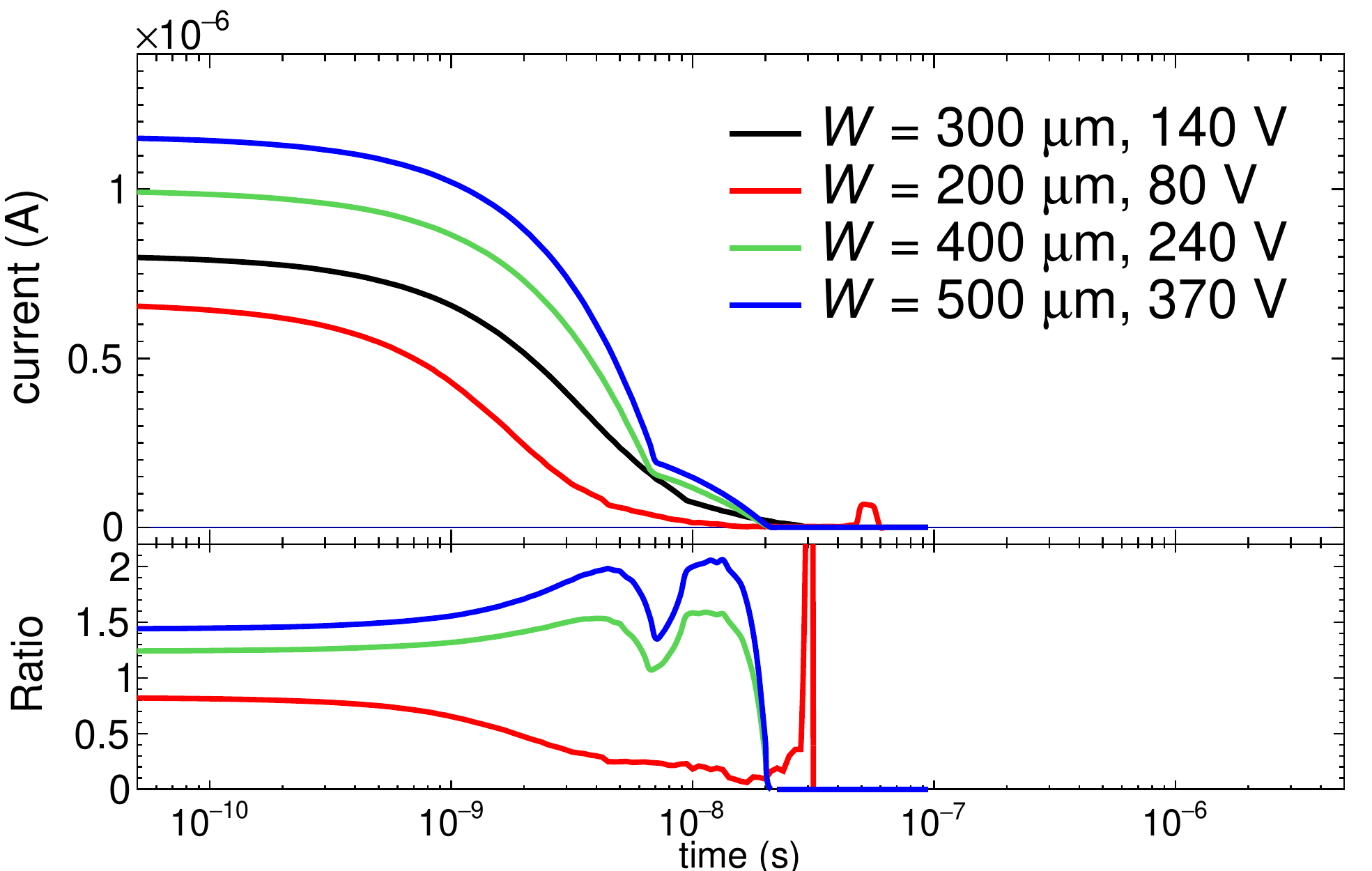}
\caption{(Color online) Total currents by varying the depth of the detector ($W$) and the reverse-bias voltage in Fast Silicon Device Simulation. The reverse-bias voltage is determined as a minimum value to makes the volume to be a whole depleted region. 
The bottom panel indicates the ratio of the current with respect to that of \unit[$W = 300$]{$\mu$m} and \unit[140]{V}.}
\label{fig:depthdependent}
\end{figure}

Figure~\ref{fig:depthdependent} shows the variation of current signals when the depth of the device changes from 200 to \unit[500]{$\mu$m}. The corresponding reverse-bias voltage is set to the minimum value that makes the whole volume of the device into a depleted region of the p-n junction. It is clear that the magnitudes and shapes of the current vary with respect to the depth of the device and the corresponding reverse-bias voltage. Also, a tail of the current signal originated by the hole current can be seen clearly around $t = 10^{-8}\,\mathrm{s}$ as the depth of the device increases.   

\section{Conclusion}
The LAMPS Collaboration at RAON~\cite{RAON} designs a detector sensitive to energy measurement for the study of low-energy heavy-ion physics with greatly enhanced isotopic identification capabilities. We developed an open-source 3-D application named Fast Silicon Device Simulation that helps us research geometric and physical properties of the silicon detector for the low energy LAMPS experiment. The simulation method calculates the electric potential and field fast with input parameters such as geometric characteristics, physical values and electrical boundary conditions of the pad detector through the GUI. 

An important output of the Fast Silicon Device Simulation is the magnitude and shape of the induced current generated by charged particles passing through the device. The current signal by Fast Silicon Device Simulation reproduces the results produced by Silvaco TCAD~\cite{SilvacoAtlas} and Garfield++~\cite{Garfieldpp} simulations qualitatively and quantitatively. The simulation results of Fast Silicon Device Simulation are more analyzed by changing variables such as the magnitude of the incident angle of the particles and the depth of the device. This simulation method can be used to study the optimized design and physical properties of the silicon detector at the low energy LAMPS experiment.  

\section{Acknowledgements}
This work was supported by the National Research Foundation of Korea (NRF) grant funded by the Korea government(MSIT) (No. 2018R1A5A1025563).

\newpage

\bibliographystyle{plain}   
\bibliography{bib.bib}

\end{document}